\documentclass[prb,twocolumn,aps,showpacs]{revtex4}
\usepackage{graphicx}
\usepackage{dcolumn}
\usepackage{bm}

\begin{document}

\preprint{APS/123-QED}
\title{D'yakonov-Perel' spin relaxation in InSb/AlInSb quantum wells}
\author{Jun Li}
\author{Kai Chang}
\email{kchang@red.semi.ac.cn}
\affiliation{SKLSM, Institute of Semiconductors, Chinese Academy of Sciences, P. O. Box
912, Beijing 100083, China}
\author{F. M. Peeters}
\affiliation{Department of Physics, University of Antwerp, Groenenborgerlaan 171, B-2020
Antwerpen, Belgium}
\date{\today}

\begin{abstract}
We investigate theoretically the D'yakonov-Perel' spin relaxation
time by solving the eight-band Kane model and Poisson equation
self-consistently. Our results show distinct behavior with the
single-band model due to the anomalous spin-orbit interactions in
narrow band-gap semiconductors, and agree well with the experiment
values reported in recent experiment (K. L. Litvinenko, et al., New
J. Phys. \textbf{8}, 49 (2006)). We find a strong resonant
enhancement of the spin relaxation time appears for spin align along
[$1\bar{1}0$] at a certain electron density at 4 K. This resonant
peak is smeared out with increasing the temperature.
\end{abstract}

\pacs{72.25.Rb, 71.28.+d, 71.70.Ej.} \maketitle

Spin relaxation time (SRT) is very important for the coherent
manipulation of electron spin and applications in spintronics
devices. There are four different spin relaxation
mechanisms\cite{SpintronicsRMD}, i.e., the D'yakonov-Perel'
(DP)\cite{DP}, Elliott-Yafet (EY)\cite{EY}, Bir-Aronov-Pikus
\cite{BAP} and hyperfine interaction\cite{HPF} mechanism. Among
them, the DP mechanism is found to be the dominating spin relaxation
mechanism in zincblende semiconductor structures over a wide range
of temperature\cite{OpOt}. According to the DP theory, the electrons
lose their initial spin orientation due to a momentum-dependent
effective magnetic field that changes its orientation frequently
which is caused by random impurity scattering. Therefore the
momentum-dependent effective magnetic field is the key factor to
determine the spin relaxation time, and it is induced by two types
of spin-orbit interactions (SOIs) in structures without inversion
symmetry. i.e., the Rashba SOI (RSOI) arising from structure
inversion asymmetry and the Dresselhaus SOI (DSOI) caused by bulk
inversion asymmetry. In conventional semiconductor quantum
structures, the DP theory based on the single-band model with linear
momentum-dependent SOIs have been demonstrated to agree well with
the experiments, e.g., GaAs/AlGaAs quantum well
(QW)\cite{GaAsQWSRT}, and InGaAs/InP QW\cite{InGaAsSRT}. Recently,
the spin relaxation time in narrow band-gap semiconductor
InSb/AlInSb QW also attracted much interest\cite
{NJPexp,JAPexp,MurdinSRT}, because of its unusual properties for
spintronics devices, e.g., small electron effective mass, strong
spin-orbit coupling, large effective Land\'{e} g factor. However,
the SOIs in narrow band-gap semiconductor is quite different from
the single-band model with linear momentum-dependent SOIs. For
example, the Rashba spin-splitting exhibits a nonlinear behavior
while the kinetic energy of the electron is comparable to the
band-gap\cite{NonliearRSS}. Therefore a detailed theoretical
investigation beyond the single band model for the DP spin
relaxation time is necessary in the case of narrow band-gap
InSb/AlInSb QW for the potential spintronics device and basic
physics.

In this work, we investigate theoretically the DP spin relaxation
time and its dependencies on the temperature, electron density and
the thickness of the InSb/AlInSb QW. The effective magnetic field
and the spin relaxation time is calculated based on the
self-consistent solution of the eight-band Kane Hamiltonian and the
Poisson equation. We find that the effective magnetic field obtained
from the eight-band model deviates strongly from that obtained by
the single-band model with the momentum-linear SOI. We show that the
eight-band model results are in good agreement with experiment
values without introducing any fitting parameter. We find a strong
anisotropic SRT: a strong resonant peak of SRT for spins aligned
along the $[1\bar{1}0]$ direction can be seen by tuning the electron
density since electron at T = 4 K. But this peak is gradually
smeared out with increasing temperature. Our results could be
helpful to observe new physical phenomenon, e.g., the intrinsic spin
Hall effect and persistent spin helix\cite{SpinHelix} in such narrow
band-gap InSb/AlInSb QWs.

\begin{figure}[t]
\includegraphics[width=\columnwidth] {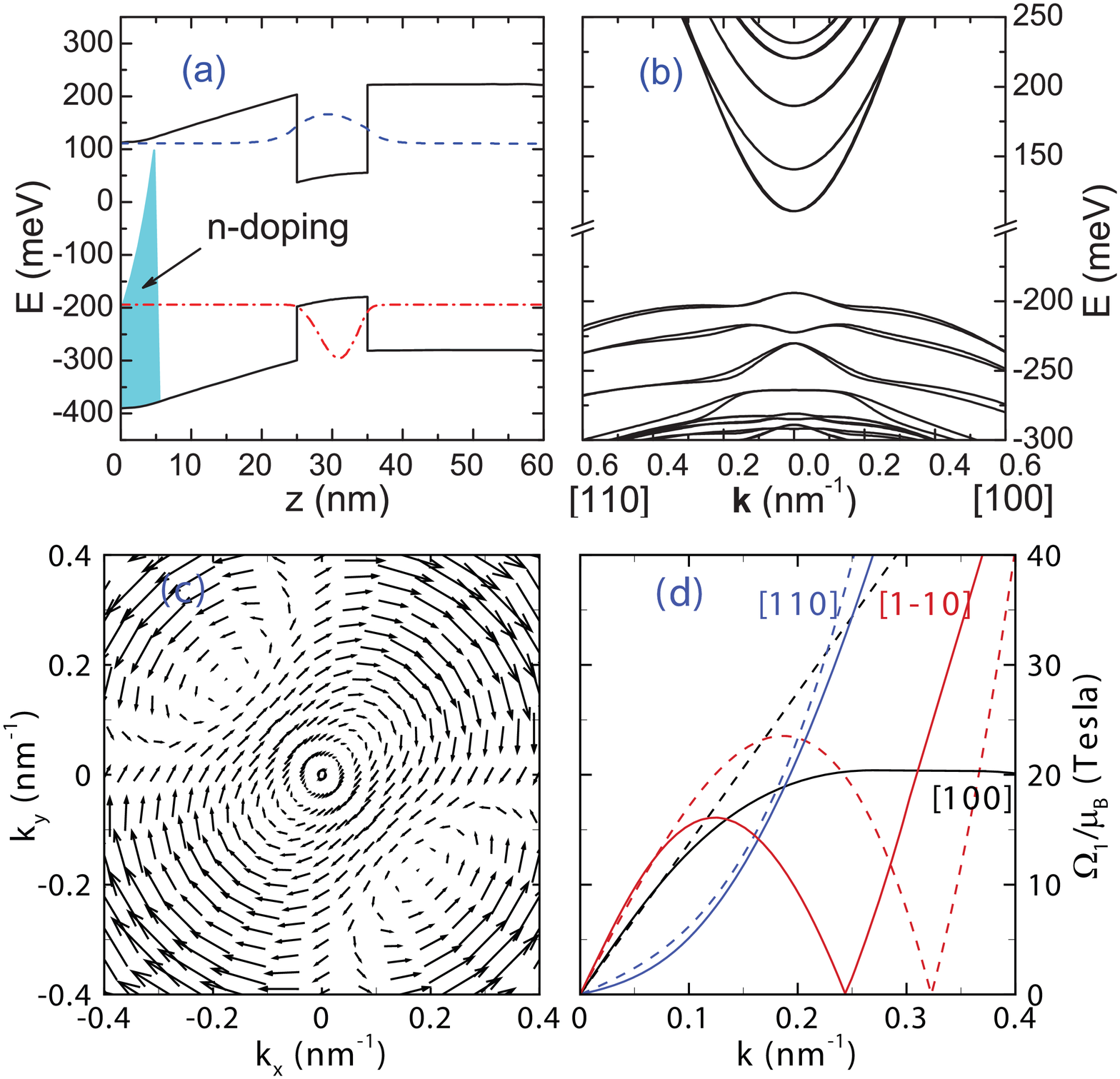}
\caption{(a) The calculated band alignment and electron (hole) probability
distribution. The shading area describes the doping profile. The sheet
carrier density is $n_{e}=5.2\times 10^{11}$ cm$^{-2}$. (b) The
self-consistently calculated energy dispersions of electron. (c) The vectors
of effective magnetic field $\boldsymbol{\Omega _{1}}(\boldsymbol{k})$ on $%
k_{x}$-$k_{y}$ plane (d) The magnitude of the effective magnetic field $%
\Omega _{1}(\boldsymbol{k})$ for $\boldsymbol{k}$ along [100](black solid
line) [110] (blue dashed line) and [$1\bar{1}0$] (red dash-dotted line) of
an asymmetrically n-doped InSb/Al$_{0.15}$In$_{0.85}$Sb QW at T = 4 K. }
\label{fig:fig1}
\end{figure}

We consider an asymmetric n-doped InSb/AlInSb QW grown along the $[001]$
crystallographic direction (see Fig. \ref{fig:fig1} (a)). The n-doping layer
is assumed to be located 20 nm on the left-side of the InSb well and with an
exponentially decaying profile. We extend the previous theory (see Refs. %
\onlinecite{Averkiev} and \onlinecite{Kainz}) to the framework of
the eight-band model by changing all operators in two-band model to
the eight-band model \cite{SHE, SpinStates}, the DP spin relaxation
time $\tau _{_{\alpha }}\,$($\alpha =+,-,z$, representing the spin
relaxation time for the spin of the injected electrons oriented
along $[110]$, $[1\overline{1}0]$, $[001]$, respectively) can be
written as

\begin{equation}
\frac{1}{\tau _{\alpha }}=4\frac{\tau _{tr}}{\hbar ^{2}}\frac{\xi
_{\alpha }^{\nu }}{\zeta ^{0}}\frac{\zeta ^{\nu}}{\zeta ^{\nu +1}},
\label{eq:SRT}
\end{equation}

with

\begin{equation}
\xi _{\alpha }^{\nu}=\sum_{s}\int_{0}^{\infty }d\boldsymbol{k}\Gamma
_{s,\alpha}(\boldsymbol{k})[E_{s}^{r}(\boldsymbol{k})]^{\nu}\Delta
F_{s,+}\left(E_{F,}\boldsymbol{k}\right) ,
\end{equation}

and

\begin{equation}
\zeta ^{\nu}=\sum_{s}\int_{0}^{\infty }d\boldsymbol{k}[E_{s}^{r}(\boldsymbol{k}%
)]^{\nu}\Delta F_{s,+}\left( E_{F,}\boldsymbol{k}\right) .
\end{equation}%
Here, $E_{s}^{r}(\boldsymbol{k})\equiv E_{s}(\boldsymbol{k})-E_{s}(0)$ is
the kinetic energy of an electron in the $s$-th subband, $\Delta
F_{s,+}\left( E_{F,}\boldsymbol{k}\right) \equiv F_{s,+}\left( E_{F,}%
\boldsymbol{k}\right) -$ $F_{s,-}\left( E_{F,}\boldsymbol{k}\right) $ is the
Fermi distribution difference between the spin-up and spin-down subband, $%
\tau _{tr}$ is the transport relaxation time and $\Gamma _{s,\alpha }(%
\boldsymbol{k})$ ($\alpha =+,-,z$) is the spin relaxation rates

\begin{equation}
\Gamma _{s,+}(\boldsymbol{k})=\Lambda _{xx}+\Lambda _{yy}-\Lambda
_{xy}-\Lambda _{yx},  \label{eq:Gamma1}
\end{equation}

\begin{equation}
\Gamma _{s,-}(\boldsymbol{k})=\Lambda _{xx}+\Lambda _{yy}+\Lambda
_{xy}+\Lambda _{yx},  \label{eq:Gamma2}
\end{equation}

\begin{equation}
\Gamma _{s,z}(\boldsymbol{k})=\Lambda _{zz}
\end{equation}%
with

\begin{equation}
\Lambda _{ij}=4\sum_{n=-\infty }^{\infty }\left[ \sum_{l}\Omega
_{s,l}^{-n}\Omega _{s,l}^{s,n}\delta _{ij}-\Omega _{s,j}^{-n}\Omega
_{s,i}^{n}\right] \eta ^{n,\nu}  \label{eq:Lammda}
\end{equation}

\begin{equation}
\eta ^{n,\nu}=\int_{0}^{2\pi }\frac{1-\cos \theta }{\sin ^{2\nu}(\theta /2)}%
/\int_{0}^{2\pi }\frac{1-\cos (n\theta )}{\sin ^{2\nu}(\theta /2)},
\label{eq:eta}
\end{equation}

\begin{equation}
\Omega _{s,i}^{n}(k)=\int_{0}^{2\pi }\frac{d\varphi _{k}}{2\pi }\Omega
_{s,i}(\boldsymbol{k})e^{-in\varphi _{k}}.
\end{equation}%
$\boldsymbol{\Omega }_{s,i}(\boldsymbol{k})$ ($i=x,y$) is the
components of the in-plane effective magnetic field of the $s$-th
subband. $\boldsymbol{\Omega }_{s,i}( \boldsymbol{k})$ can be
obtained by ascribing the spin-splitting induced by space inversion
asymmetry to the Zeeman splitting caused by the effective magnetic
field. Therefore, by using the eight-band Zeeman term\cite{Kainz}
$H_{z}=\mu _{B}\boldsymbol{B\cdot \Sigma }$, $\boldsymbol{\Omega
}_{s,i}(\boldsymbol{k})$ can be written as
\begin{equation}
\Omega _{s,i}(\boldsymbol{k})\equiv \mu _{B}B_{i}=\frac{S_{s,i}(\boldsymbol{k%
})\Delta E_{s}(\boldsymbol{k})}{2\sqrt{S_{s,x}(\boldsymbol{k})^{2}+S_{s,y}(%
\boldsymbol{k})^{2}}},~  \label{eq:EMF}
\end{equation}

\begin{equation}
S_{s.i}(\boldsymbol{k})=\langle \psi _{s,+}(\boldsymbol{k})|\Sigma _{i}|\psi
_{s,+}(\boldsymbol{k})\rangle -\langle \psi _{s,-}(\boldsymbol{k})|\Sigma
_{i}|\psi _{s,-}(\boldsymbol{k})\rangle .
\end{equation}
$\Sigma _{i}$($i=x,y$) is the components of the eight-band effective
spin matrices which can be found in Ref. \onlinecite{SpinStates} and
\onlinecite{SOIRWinkler}. $\Delta E_{s}(%
\boldsymbol{k})\equiv E_{s,+}(\boldsymbol{k})-E_{s,-}(\boldsymbol{k})$ is
the spin-splitting of the $s$-th subband. The eigen-energy $E_{s,\pm }(%
\boldsymbol{k})$ and eigen-states $\left\vert \psi _{s,+}(\boldsymbol{k}%
)\right\rangle $ can be numerically obtained by solving the
eight-band Kane Hamiltonian and the Poisson equation
self-consistently\cite{SpinStates}. Through this approach, the
non-parabolic effect and the anomalous behavior of SOIs in
narrow band-gap semiconductors can be taken into account. In Eq. (\ref{eq:SRT}%
)-(\ref{eq:eta}), $\nu $ is a constant characterizing the relation
of momentum scattering time on the electron kinetic energy ($\tau
_{p}(\boldsymbol{k})\propto \lbrack
E_{s}^{r}(\boldsymbol{k})]^{\nu}$). For acoustic phonon and screened
ionized impurities scattering (type I), $\nu =0 $, for polar optical
phonon scattering (type II), $\nu =1$, for weekly screened ionized
impurities (type III), $\nu =2$\cite{Kainz}.

In Figs. \ref{fig:fig1}(a) and (b) we show the calculated potential profile,
the electron probability and the energy dispersion of a 10 nm InSb/Al$%
_{0.15} $In$_{0.85}$Sb QW at T = 4 K including the effect of
built-in electric field caused by the charge redistribution. All the
Kane parameters of the materials used in our calculation are taken
from Ref. \onlinecite{Parameters}, and the ratio of the conduction
band offset and the conduction band offset is taken as
62\%:38\%\cite{Bandoffset}. The bulk inversion asymmetry of
zincblende crystal is introduced by the $B$ parameter in the
eight-band Kane
Hamiltonian\cite{KaneHamitonian}, which is taken to be $B=31.4~eV\cdot {%
\mathring{A}}$\cite{MurdinSRT}. Besides, we should notice that the
temperature dependence of the bandgap (Varshini
relation)\cite{Varshni} is more pronounced in a narrow bandgap
semiconductor than that in a wide bandgap semiconductor, for
instance, the bulk bandgap of InSb, of which the bulk band gap is
0.235 eV at 4 K, and 0.174 eV at 300 K, i.e., up to a 26\% variation
of the bandgap with increasing temperature. Because the SOI and
spin-splitting is intimately related to the conduction-valence band
coupling, the decreasing of band gap could lead to an enhancement of
the SOI and results in the increasing of electron subband
spin-splitting and the effective magnetic field (about 11\%).

Fig. \ref{fig:fig1}(c) shows the effective magnetic field as a
function of the in-plane momentum. The effective magnetic field of a
$(001)$-grown InSb/AlInSb QW always lie in the QW plane. Due to the
interplay of RSOI and DSOI, the effective magnetic field exhibits a
$C_{2v}$ symmetry. Notice that the effective magnetic field is
enhanced or weakened when $\boldsymbol{k}$ along the $[110]$ or
$[1\bar{1}0]$ directions. Fig. \ref{fig:fig1}(d) shows the self
consistent eight-band modeling for the magnitude of the effective magnetic field for $%
\boldsymbol{k}$ along $[100],[110],[1\bar{1}0]$ (see the solid curves).
Along $[1\bar{1}0]$ crystallographic direction, the effective magnetic field pointing along $[%
\bar{1}\bar{1}0]$ vanishes at a certain Fermi wavevector $\boldsymbol{k}$,
which makes the spin lifetime $\tau _{_{-}}$ become very long. In Fig. \ref%
{fig:fig1} (d) we compare these results with those from the single-band
model with $\boldsymbol{k}$-linear SOI. One can see clearly that the
deviation of effective magnetic field is very large (up to 42\%) at $\boldsymbol{k}=0.2$ nm$^{-1}$%
($[100]$). This deviation comes from the weakening of the interband coupling
as the electron kinetic energy becomes comparable to the bandgap. Therefore
the single-band model may not be good enough to describe the strong SOIs in
such narrow bandgap QWs.

\begin{figure}[t]
\includegraphics[width=\columnwidth] {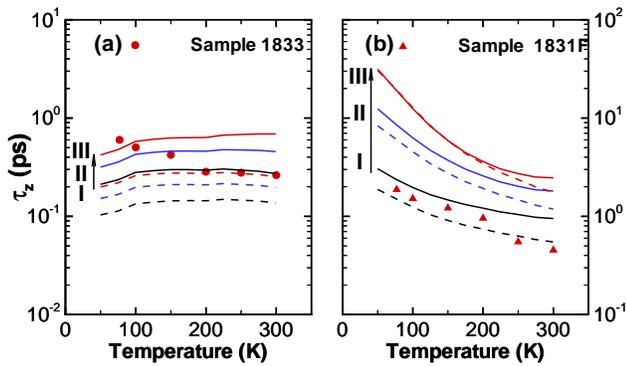}
\caption{Calculated spin relaxation time $\protect\tau _{z}$ versus
temperature compare to the experimental results reported in Ref.
\onlinecite{NJPexp}. (a) Sample 1833 (asymmetrically n-doped 20 nm InSb/Al$%
_{0.15}$In$_{0.85}$Sb QW) (b) Sample 1831F (uniformly n-doped 20 nm InSb/Al$%
_{0.15}$In$_{0.85}$Sb QW). The black, blue, red lines represent the
eight-band numerical results for type I ($\protect\nu =0$), type II ($%
\protect\nu =1$), and type III ($\protect\nu =2$) momentum scattering
mechanism. The dashed lines with the same colors represent the results from
the single-band model with linear SOI.}
\label{fig:fig2}
\end{figure}

In Fig. \ref{fig:fig2} we compare the numerical results of SRT $\tau
_{z}$ with the experimental measurement reported in Ref.
\onlinecite{NJPexp}. The transport momentum relaxation time $\tau
_{tr}$ used in our calculation are obtained from the measured Hall
mobility (by $\tau _{tr}=m^{\ast }\mu _{Hall}/\left( e r
_{Hall}\right) $\cite{Kainz}). The electron density is
assumed to increase linearly from $3.6\times 10^{11}$ cm$^{-2}$ to $%
5.3\times 10^{11}$ cm$^{-2}$ for sample 1833 and from $5.7\times 10^{11}$ cm$%
^{-2}$ to $7.3\times 10^{11}$ cm$^{-2}$ for sample 1831F when
temperature increase from 77 K to 300 K. Considering different
momentum relaxation mechanisms, our eight-band numerical results
agree quite well with the experiment values without having to
introduce any fitting parameter. From panel (a), we can see the
weekly screened impurity scattering ($\nu =2$) and polar phonon
scattering ($\nu =1$) dominate at $T<150$ K and the ionized impurity
scattering dominates at $T>150$ K. Noticed that in heavily doped
semiconductor samples, the dominant momentum scattering mechanisms
varied through neutral (weekly screened) impurities scattering,
acoustic and polar phonon scattering, and ionized impurity
scattering with increasing temperature\cite{Ridley}, therefore our
results are reasonable and also consistent with the previous
work\cite{Kainz,Puller}. One can see that, due to the overestimate
of SOI strength by the linear SOI model, the single-band model will
underestimate the SRT compared to the eight-band model and doesn't
agree with the measured SRT. For the uniformly doped sample 1831F,
our calculated DP SRT is larger than the measured value. The
discrepancy between the calculated DP SRT from the eight-band model
with measured value is reasonable because the SRT induced by the EY
mechanism could be comparable to the DP SRT\cite{NJPexp}. The
calculated SRT from the single-band model in this symmetric doped
sample is similar with that from the eight-band model. This is
because the cubic DSOI term (the RSOI is absent due to symmetric
doping) in this sample may play a dominant role.
\begin{figure}[t]
\includegraphics[width=\columnwidth] {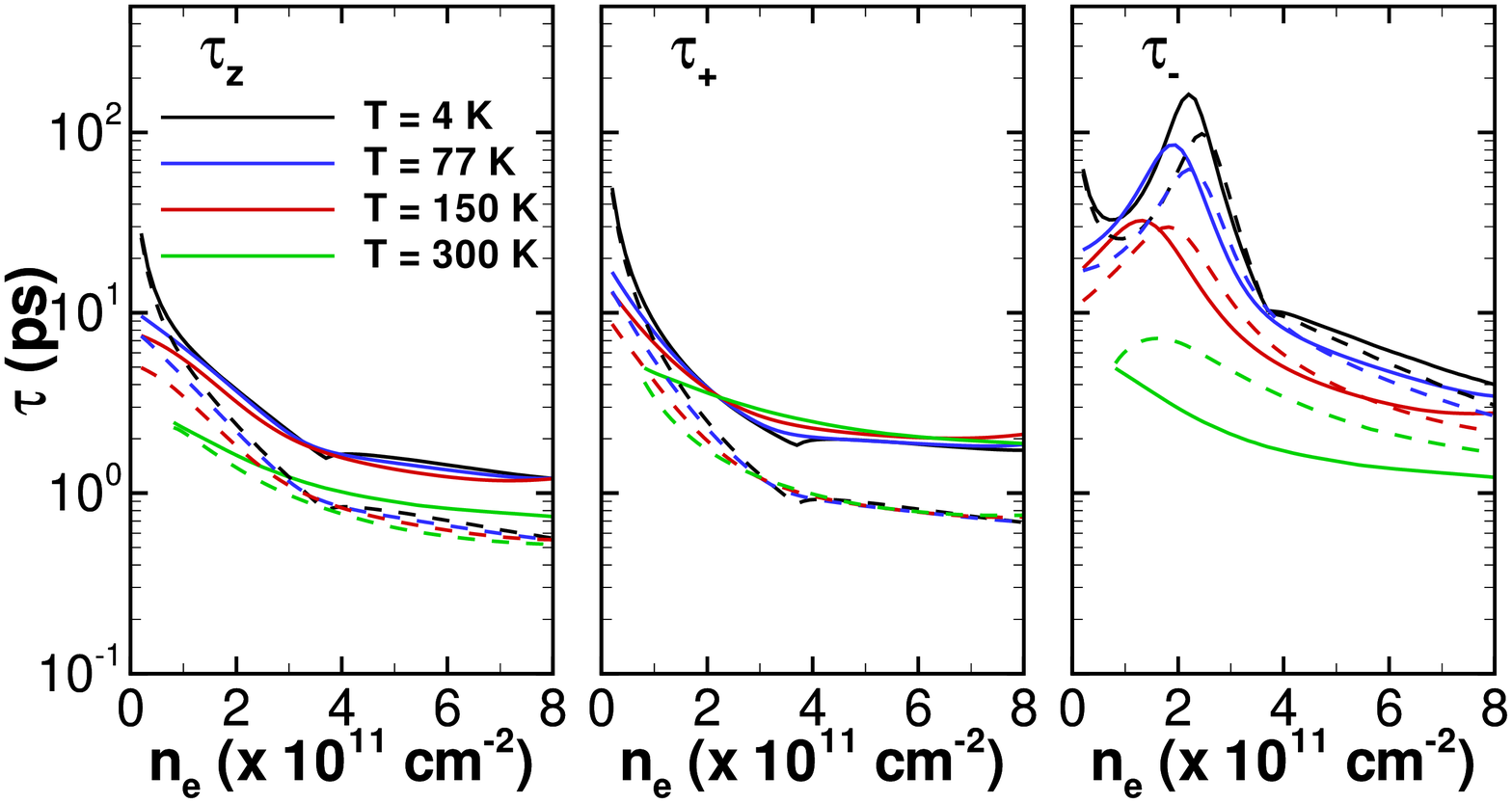}
\caption{Spin relaxation times $\protect\tau _{z}$, $\protect\tau _{+}$ and $%
\protect\tau _{-}$ as a function of electron density $n_{e}$ in a 10 nm
asymmetric n-doped InSb/Al$_{0.15}$In$_{0.85}$Sb QW for different
temperature. The dashed lines with the same colors represent the results of
single-band model with linear SOI.}
\label{fig:fig3}
\end{figure}

In Fig. \ref{fig:fig3} we calculate the spin relaxation times $\tau _{z}$, $\tau _{+}$ and $%
\tau _{-}$ as a function of electron density $n_{e}$ in a 10 nm n-doped
InSb/Al$_{0.15}$In$_{0.85}$Sb QW for different temperatures. As shown in the
figure, the SRTs $\tau _{z}$, $\tau _{+}$ decrease with increasing electron
density due to the enhancement of SOIs with increasing the Fermi wavevector.
When temperature increases, the $\tau _{z}$ and $\tau _{-}$ are suppressed
strongly, but $\tau _{+}$ is not very sensitive to temperature. The giant
spin relaxation anisotropy\cite{Averkiev, Kainz, MurdinSRT} is also
demonstrated in InSb/AlInSb QW: the SRT of the $[1\bar{1}0]$-oriented spins
shows a resonant peak at a certain electron density at low temperature.
However, the resonant peak is very sensitive to temperature. As temperature
increases from 4K to 300K, the peak is gradually smeared out, which is due
to the bluring of the Fermi surface with increasing temperature. The dashed
lines show the results of the single-band model. Consistent with the former
discussion, the results of the single-band model is smaller than that of the
eight-band model but at small $n_{e}$.

\begin{figure}[t]
\includegraphics[width=\columnwidth] {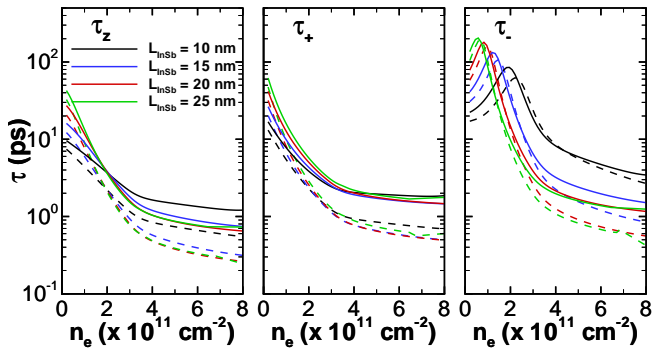}
\caption{Spin relaxation times $\protect\tau _{z}$, $\protect\tau _{+}$ and $%
\protect\tau _{-}$ as a function of carrier density in asymmetric n-doped
InSb/Al$_{0.15}$In$_{0.85}$Sb QWs. The black solid, blue dashed, red dashdot
and green dotted lines represent the numerical results of $L_{InSb}=10,15,20$
and $25$ nm. The dashed lines with the same colors represent the results of
the single-band model with linear SOI.}
\label{fig:fig4}
\end{figure}

In Fig. \ref{fig:fig4} we exhibit the SRTs $\tau _{z}$, $\tau _{+}$ and $%
\tau _{-}$ as a function of electron density in a 10 nm n-doped InSb/Al$%
_{0.15}$In$_{0.85}$Sb QW for different thicknesses of InSb well. The SRTs
increase with increasing thickness of the InSb well when the electron
density $n_{e}$ is small, but has an opposite trend when the electron
density $n_{e}$ are larger than a certain value. The resonant peak value of $%
\tau _{-}$ becomes larger and the corresponding $n_{e}$ becomes
smaller with increasing thickness of the QW. The results of the
single-band model is smaller than that of the eight-band model
except at small $n_{e}$.

In summary, we investigated theoretically the SRT in InSb/AlInSb QW
beyond the single-band model. Our results are obtained within the
eight-band model and agree very well with the measured SRTs, while
the SRT obtained from the single-band model with linear
momentum-dependent SOIs deviates strongly from that of the
eight-band model due to the strong interband-coupling in narrow
bandgap QWs. We also demonstrate that the SRT along
$[1\bar{1}0]$-direction shows a resonant peak at a certain electron
density, i.e., very long SRT. The resonant peak will be smeared out
with increasing temperature.

\begin{acknowledgments}
This work is supported by the NSFC Grant No. 60525405 and 10874175, and the
bilateral program between China and Belgium, and the Belgian Science Policy
(IAP).
\end{acknowledgments}

\end{document}